\documentclass[12pt,preprint]{aastex}

\slugcomment{Accepted by the Astrophysical Journal}

\shorttitle{Upper Limits on Pulsed Radio Emission from XTE~J0103$-$728}

\shortauthors{Crawford et al.}

\begin{document}

\title{Upper Limits on Pulsed Radio Emission from the 6.85 s X-ray Pulsar
XTE~J0103$-$728 in the Small Magellanic Cloud}

\author{Fronefield Crawford\altaffilmark{1}, Duncan
R. Lorimer\altaffilmark{2}, Brian M. Devour\altaffilmark{1}, Brian
P. Takacs\altaffilmark{1}, \& Vladislav I. Kondratiev\altaffilmark{2,3,4}}
 
\altaffiltext{1}{Department of Physics and Astronomy, Franklin and
Marshall College, P.O. Box 3003, Lancaster, PA 17604, USA; email:
fcrawfor@fandm.edu}
 
\altaffiltext{2}{Department of Physics, West Virginia University,
Morgantown, WV 26506, USA}

\altaffiltext{3}{National Radio Astronomy Observatory, P.O. Box 2,
Green Bank, WV 24944, USA}

\altaffiltext{4}{Astro Space Center of the Lebedev Physical Institute,
Profsoyuznaya str. 84/32, Moscow 117997, Russia}

\begin{abstract}
X-ray pulsations with a 6.85 s period were recently detected in the
Small Magellanic Cloud (SMC) and were subsequently identified as
originating from the Be/X-ray binary system XTE~J0103$-$728. The
recent localization of the source of the X-ray emission has made a
targeted search for radio pulsations from this source possible. The
detection of pulsed radio emission from XTE~J0103$-$728 would make it
only the second system after PSR B1259$-$63 that is both a Be/X-ray
binary and a radio pulsar.  We observed XTE~J0103$-$728 in Feb 2008
with the Parkes 64-m radio telescope soon after the identification of
the source of X-ray pulsations was reported in order to search for
corresponding radio pulsations.  We used a continuous 6.4 hour
observation with a 256 MHz bandwidth centered at 1390 MHz using the
center beam of the Parkes multibeam receiver. In the subsequent data
analysis, which included a folding search, a Fourier search, a
fast-folding algorithm search, and a single-pulse search, no pulsed
signals were found for trial dispersion measures (DMs) between 0 and
800 pc cm$^{-3}$. This DM range easily encompasses the expected values
for sources in the SMC.  We place an upper limit of $\sim 45$ mJy
kpc$^{2}$ on the luminosity of periodic radio emission from
XTE~J0103$-$728 at the epoch of our observation, and we compare this
limit to a range of luminosities measured for PSR B1259$-$63, the only
Be/X-ray binary currently known to emit radio pulses. We also compare
our limit to the radio luminosities of neutron stars having similarly
long spin periods to XTE~J0103$-$728.  Since the radio pulses from PSR
B1259$-$63 are eclipsed and undetectable during the portion of the
orbit near periastron, repeated additional radio search observations
of XTE~J0103$-$728 may be valuable if it is undergoing similar 
eclipsing and if such observations are able to sample the orbital
phase of this system well.
\end{abstract}


\section{Introduction} 

Neutron stars appear as detectable sources in a variety of forms.  Of
the almost 1800 pulsars that are currently known in the ATNF pulsar
catalog
\citep{mht+05}\footnote{http://www.atnf.csiro.au/research/pulsar/psrcat/},
the vast majority are detectable only as radio sources. However, some
of these neutron stars are visible as both pulsed X-ray and radio
sources, and some pulsars emit pulsed radiation only at high energies
(X-rays or gamma-rays). A small number of pulsars are also visible at
optical wavelengths.  Apart from these generally persistent sources,
transient radio emitters may also exist in large numbers
\citep{mll+06}, and some neutron stars are known to be transient X-ray
sources. For instance, radio pulsations from Anomalous X-ray Pulsars
(AXPs) undergoing X-ray outbursts have recently been detected
\citep{crh+06,crh+07}, while radio emission from several other AXPs
has not \citep{bri+06,chk07}. These results have been used in part to
study the physics of this class of highly magnetized neutron stars
(``magnetars'').  The absence of detectable radio emission in searches
of central compact objects \citep{psg+02,d08}, nearby thermal X-ray
pulsars having long periods (X-ray dim isolated neutron stars; XDINs)
\citep[e.g.,][]{bj99,j03,kbp+08,kml+08}, and the magnetar-like (i.e.,
young, having a strong magnetic field, and exhibiting X-ray outbursts)
X-ray pulsar PSR J1846$-$0256 \citep{akl+08} suggests that the
emission physics of neutron stars and the connection between typical
radio pulsars and these other kinds of objects are in many ways not
well understood.  Like some of these X-ray systems, radio pulsars can
have long spin periods, but these radio pulsars usually have surface magnetic
field strengths that are more typical for the radio pulsar population.
PSR J2144$-$3933 is such an example: it is detectable only as a
low-luminosity radio source, has a very long spin period (8.5 s), and
has a very small pulsed duty cycle \citep{ym99}, which suggests that
these kinds of systems may be more common than is observed but are
easily missed.  In short, the study of neutron star systems at
multiple wavelengths can provide different windows into the physics of
these objects, so a multiwavelength approach is fruitful whenever
possible.

Pulsars that are members of binary systems are particularly useful as
physical probes, allowing an investigation of the companion star, the
pulsar itself, the dynamics and evolution of the system, and theories
of gravity.  An extensive review is presented by \citet{l08}. A particular
class of high-mass X-ray binary (HMXB) consists of a neutron star
orbiting a Be star, which is an early-type non-supergiant star with
observable emission lines from the material in its circumstellar disk
(see Slettebak 1988\nocite{s88} for a review). Only one Be/X-ray
binary is also known to be a radio pulsar: PSR B1259$-$63
\citep{jml+92}.  PSR B1259$-$63 has been extensively studied as both a
radio pulsar and transient X-ray source since its discovery almost 20
years ago, particularly at or near its periastron passages, which
occur every 3.4 years \citep[e.g.,][]{ktn+95,jml+96,jmm+99,jbw+05}.

X-ray pulsations with a period of 6.8482(7) s\footnote{The figure in
parentheses represents the uncertainty in the last digit quoted.} were
first detected by RXTE in May 2003 from an unlocalized and
unidentified source in the Small Magellanic Cloud (SMC)
\citep{cmm+03}.  More recently, \citet{hp08} announced the detection
of an X-ray transient source in an XMM-Newton observation from Oct
2006 in which these pulsations were again clearly detected. In this
latter observation, the pulse period was constrained to be 6.85401(1)
s, and the source was localized to the position $\alpha$ = 01:02:53.9,
$\delta$ = $-$72:44:34.6 (J2000), with a position uncertainty of $\sim
1''$. This tight position constraint unambiguously identified the
source as the Be/X-ray binary XTE~J0103$-$728, one of the many HMXBs
of this type in the SMC \citep{hp08}. Given the very accurate position
of the source as determined by this X-ray observation, a deep radio
search for radio pulsations became feasible.  The detection and study
of any pulsed radio emission from this source would be interesting
given the as-yet unclear connections between radio emission and X-ray
activity in the classes of neutron stars described above, plus the
physical insight to be gained from finding and studying another
Be/X-ray binary radio pulsar.

\section{Observations and Analysis} 

On 1 Feb 2008, we observed the source XTE~J0103$-$728 continuously for
23106 s (6.4 hours) with the Parkes 64-m telescope in Parkes,
Australia. Data were taken with the center beam of the multibeam
receiver (the most sensitive of the 13 receiver beams) at a center
frequency of 1390 MHz.  One reason this relatively high observing
frequency was used is that PSR B1259$-$63, which is a possible
comparison system, is known to be particularly weak at low frequencies
($\sim 400$ MHz), possibly owing to an unusual spectral index
\citep{jml+94}.  A bandwidth of 256 MHz was used in our observation,
split into 512 contiguous 0.5 MHz frequency channels.  Dual linear
polarizations were summed, and the frequency channels were one-bit
sampled every 0.5 ms. This easily preserved the necessary sampling
rate required to detect 6.85 s pulsations. The data were recorded on
DLT tape and transferred from the observatory to a Beowulf cluster for
processing.  The SMC was previously surveyed for radio pulsars at this
frequency \citep{mfl+06}, but that survey used a shorter integration
time, and XTE~J0103$-$728 was not near the center of any of the survey
beams.  These factors made that survey less sensitive to radio pulses
from XTE~J0103$-$728 by about a factor of 3 compared to the search
reported here.

The data from the Parkes observation were first checked for the
presence of radio frequency interference (RFI). Interference signals
above the 4$\sigma$ threshold level were excised from the data, which
removed $\sim 8$\% of the time bins and frequency channels in the raw
data prior to analysis. The data were then searched in several ways to
look for radio pulsations, as described below.

\subsection{Folding Search}

In the first analysis, the data were split into sixteen 16-MHz
subbands, and each subband was dedispersed at an assumed dispersion
measure (DM) of 100 pc cm$^{-3}$. This is within the range of expected
DMs for pulsars in the SMC \citep{mfl+06}.  A range of DMs was tried
by shifting the subbands with respect to each other in phase, then
summing the subbands in each case. DMs from 0 to 5000 pc cm$^{-3}$
were searched in this way.  A barycentric folding period was
determined using the measured periods from the two previous X-ray
detections of XTE J0103$-$728. Using the reported periods and their
associated uncertainties from the May 2003 \citep{cmm+03} and Oct 2006
\citep{hp08} observations, we extrapolated the period to the Feb 2008
epoch and determined the barycentric folding period to be 6.8563(3)~s.
Since the spin period of XTE~J0103$-$728 is so large, a dedispersion
error would not significantly degrade the sensitivity of our search. A
dedispersion error of 100 pc cm$^{-3}$, for instance, would lead to
only a 5 ms smear in the pulse across each subband, less than 0.1\% of
the pulse period. Even for the extreme case of a dispersion error of
5000 pc cm$^{-3}$, pulsations would probably still be detectable since
the corresponding pulse smear would only be 3.6\% of the pulse period.

Periods $\pm 40$~ms from the nominal period were searched, which
easily encompassed the estimated 0.3 ms uncertainty in the pulse
period.  In a separate analysis, the data were split into ten
segments, with each segment starting at intervals of 10\% of the data
length and consisting of 20\% of the data length.  Each segment was
separately searched for variable strength pulsations (owing to
scintillation, for example) in the manner described above. No
candidate signals were detected in any of these analyses.

We also conducted a more detailed separate search in which we
dedispersed and folded the data into pulse profiles at a range of DMs
and folding periods around the nominal period. Periods spanning $\pm
1.5$~ms from the nominal 6856.3 ms pulse period (i.e., a range of $\pm
5\sigma$) were searched in steps of 1 $\mu$s. Each of these folding
trials was conducted for DMs between 0 and 800 pc cm$^{-3}$ in steps
of 5 pc cm$^{-3}$. For each DM trial, the full 256 MHz of bandwidth
was dedispersed. A total of 480000 period and DM combinations were
tried (3000 folds per DM trial, and 160 DM trials), and for each trial
the $\chi^{2}$ significance of the folded profile was recorded. For
this number of random trials, we would expect a $\sim 25$\% likelihood
of producing at least one signal at the $5\sigma$ significance level
or higher from noise alone. We therefore chose $5\sigma$ as a
reasonable threshold to apply, and we found no folded profiles at the
$5\sigma$ significance level or higher in the search.

This second search was more sensitive to pulsations having very small
duty cycles, and we estimate the effects on the pulse width that
either folding the raw data at a period that is offset from the true
period or dedispersing the data at a DM that is offset from the true
DM would have. We compare this amount of pulse smearing to the
smearing already present in the finite frequency channels. Significant
smearing of the pulse would greatly reduce the sensitivity of the
search, so it is therefore important to ensure that the search
parameters do not allow this.  For the chosen period step size of
10$^{-6}$ s in the search, the maximum folding period offset is $5
\times 10^{-7}$ s from the true period. Folding the raw data at this
offset period would cause a pulse smear across the folded profile of
1.7 ms ($2.5 \times 10^{-4}$ cycles) over the length of the
observation. This is an order of magnitude larger than the
intra-channel dispersive smearing of 0.16 ms ($2.3 \times 10^{-5}$
cycles) for an assumed DM~$\sim 100$ pc cm$^{-3}$, but it is still
negligible for any reasonable duty cycle expected for the pulsar. The
DM step size of 5 pc cm$^{-3}$ produces a maximum DM offset from the
true DM of 2.5 pc cm$^{-3}$, which for 256 MHz of bandwidth at our
observing frequency yields a maximum dispersive smearing of 2.0 ms
($2.9 \times 10^{-4}$ cycles). Thus, any duty cycle $\ga 10^{-3}$
should not be significantly affected by any of these smearing effects.
The grid spacing of the period and DM search ensured that no
pulsations with very small duty cycles (narrow pulses) would have been
missed in the search.

\subsection{Periodicity Search}

We also conducted a periodicity search of the data, which included a
Fast Folding Algorithm (FFA) search in addition to the more common
Fourier search.  A Fourier search is particularly important if the
estimated spin period at the epoch of observation was not correct
(e.g., if there were irregularities in the spin-down behavior or
significant accelerations from binary motion that are not accounted
for in the predicted folding period). For the Fourier search, two
different packages were used:
SIGPROC\footnote{http://sigproc.sourceforge.net}
\citep[e.g.,][]{lkm+00} and
PRESTO\footnote{http://www.cv.nrao.edu/$\sim$sransom/presto}
\citep{r01,rem02}. A total of 828 DM trials ranging from 0 to 800 pc
cm$^{-3}$ were used in the search. The DM trials were spaced so that
there would be no significant pulse smearing introduced beyond the
smearing already present in the 0.5 MHz frequency channels. Each
dedispersed time series was filtered, and a 2$^{25}$ point Fast
Fourier transform was performed\footnote{For the PRESTO search, each
dedispersed time series was padded out to 2$^{26}$ points so that all
of the data would be used.}. Each resulting spectrum was harmonically
summed using 16 harmonics and searched for candidate
periodicities. For the PRESTO search, an acceleration search spanning
$\pm 100$ Fourier bins was also performed on each dedispersed time
series. This maintained full sensitivity to accelerations of up to 24
m s$^{-2}$ for the 6.856 s fundamental and 16 of its harmonics.  No
acceleration search was performed in the SIGPROC analysis. Candidates
recorded in the frequency spectrum were further checked by
dedispersing and folding the raw data at DMs and periods near the
candidate values in the manner described in the folding search
above. No promising candidates were detected in this search.

During the Fourier search, each dedispersed time series was also
searched for significant periodicities using the FFA \citep{s69}. We
used publicly available FFA search
code\footnote{http://astro.wvu.edu/projects/xdins}, described in
detail by \citet{kml+08}, which has been previously used to search for
pulsed radio emission from XDINs. No signals were detected in the FFA
periodograms above a signal-to-noise threshold of 5 for periods
between 2 and 12~s.

\subsection{Single Pulse Search}

The recent discovery of bursting radio sources subsequently identified
as rotating neutron stars (RRATs; McLaughlin et
al. 2006\nocite{mll+06}) suggests that looking for dispersed single
pulses is a powerful method by which to detect rotating neutron stars
that would otherwise not be detectable in a periodic search.  Besides
RRATs, other classes of persistent and transient radio sources are
sometimes detectable through their single pulses. For example, the AXP
XTE J1810$-$197 exhibited very bright single pulses at the time of its
radio detection \citep{crh+06}, and single pulse searches have been
used to detect radio emission from a number of pulsars, including the
Crab pulsar \citep{sr68}.  Dispersed, extremely luminous radio bursts
from cosmological sources may also be present in the data and would
only be detectable through this kind of search \citep{lbm+07}.

We searched for dispersed single pulses in the data using the same set
of DMs used for the Fourier search described above (828 DMs of
variable spacing from 0 to 800 pc cm$^{-3}$). The single pulse
analysis procedure is outlined by \citet{cm03} and uses a matched
filtering approach to detect impulses of different widths in each
dedispersed time series. A range of boxcar filter widths is used to
maximize sensitivity on different time scales. Pulse widths between
0.5 ms and 1024 ms were searched by increasing factors of two for the
boxcar filter width, and the detection threshold was determined for
each filtering window by the level of RFI present.  For pulse widths
$\ga 1$~s, the high-pass filter present in the observing system would
significantly attenuate the impulse amplitude (see discussion below),
so windows greater than $\sim 1$ s were not searched.  Owing to the
increased RFI presence as the window size increased, the
signal-to-noise (S/N) detection threshold changed from 8 to 20 for
boxcar windows greater than or equal to 4 ms.  No significant
dispersed single pulses were detected in this search.

\subsection{Limits on Radio Emission}

We used the modified radiometer equation \citep{dtw+85} to estimate
the flux density limit on periodic emission in the folding search. For
a detection S/N threshold of 5, 1-bit sampling, a gain of 0.735 K
Jy$^{-1}$ for the center beam of the multibeam receiver
\citep{mlc+01}, an assumed duty cycle of 5\%, and the observing
parameters outlined above, we estimate the 1400 MHz flux density limit
on pulsed emission from XTE J0103$-$728 to be $\sim 13 \mu$Jy.  This
estimate does not include the effect of RFI, which can quite
significantly degrade the sensitivity, especially at long periods.
The corresponding 1400 MHz luminosity limit can be calculated using $L
= S d^{2}$, where $S$ is the 1400 MHz flux density limit and $d$ is
the distance. In this definition, no geometrical factor for the
beaming is included which would change the calculated luminosity
limit. Using a distance of 60 kpc to the SMC \citep{hhh05}, we
estimate the 1400-MHz luminosity limit to be $\sim 45$ mJy kpc$^{2}$
(see Table \ref{tbl-1}).  The sensitivity limit from the FFA search is
comparable to the radiometer limit for the periodic emission.

For the single pulse search, no pulses were detected down to the flux
density limit of 0.60-0.03 Jy (depending on the S/N threshold and
window size used in the filtering; see the discussion above). These
limits correspond to 1400 MHz luminosity limits of 2150-120 Jy
kpc$^{2}$ (Table \ref{tbl-1}) using a distance of 60 kpc to the SMC
and the same luminosity definition.

The limits obtained by the modified radiometer equation are accurate
representations of the sensitivity of the folding search
\citep[see][]{dtw+85,lfl+06}. However, for the sample of pulsars that
were detected in the Parkes Multibeam Survey \citep{mlc+01}, the S/N
values of the Fourier detections from the blind Fourier search are
lower than the corresponding folded profile S/N values by about a
factor of two (on average) near the detection threshold (Crawford et
al., in prep.). This difference would apply here also since the same
observing system and a similar analysis procedure were used. We
therefore estimate the baseline sensitivity limit for the Fourier
search to be $\sim 25 \mu$Jy, roughly twice the modified radiometer
equation limit. Since we had a good estimate of the folding period for
this search, we elect to keep the radiometer limit as an accurate
estimate of the sensitivity to periodic emission for our discussion.

Another factor which must be considered in the estimate of the
sensitivity to pulsed emission (either impulsive or periodic) is the
presence of a high-pass filter in the observing system which severely
limits the sensitivity to long-period pulsars having large duty
cycles. This high-pass filter is modeled as a two-pole filter with a
time-constant of 0.9 s \citep{mlc+01}, and the effect of this filter
on long-period signals depends on the duty cycle of the pulse as it is
received at the telescope (i.e., after interstellar dispersive and
scattering effects have already affected the pulse width and
shape). In order to quantify this filtering effect on the sensitivity,
we created synthetic pulse trains in software with various periods and
duty cycles and passed these signals through a two-pole high pass
filter (in software) in order to gauge the amount of signal
attenuation and hence reduction in the detected S/N.  We found that
for duty cycles greater than $\sim 20$-30\%, the attenuation is
significant for a $\sim 6$ s pulsar (there is a reduction of $\ga
10$\% in the pulse amplitude). For duty cycles less than $\sim 10$\%,
the effect is negligible; this is due to the presence of a larger
number of high-frequency harmonics of the fundamental in the signal
that are not as significantly affected by the high-pass
filtering. These results are consistent with the simple expectation
that pulse widths greater than the 0.9 s filtering time constant would
be significantly attenuated.  This filtering effect also has relevance
for the flux density upper limits recently reported for several radio
search observations of AXPs conducted with Parkes. Both \cite{bri+06}
and \citet{chk07} searched four AXPs with periods of 6 and 11 s with
the multibeam observing system. For small assumed duty cycles (less
than 10\%), the attenuation is negligible for these periods. For our
single pulse search of XTE~J0103$-$728, impulsive signals having widths
$\ga 1$~s are attenuated by the filter, which is why our single pulse
search window range did not extend beyond 1024 ms.

\section{Discussion}

Our 1400 MHz luminosity upper limit for the periodic radio emission
from XTE~J0103$-$728 is estimated to be $\sim 45$ mJy kpc$^{2}$ (Table
\ref{tbl-1}). Below we compare this limit to the luminosities of
neutron star systems having characteristics similar to
XTE~J0103$-$728. In particular, we compare our result to PSR
B1259$-$63, the one known radio pulsar in a Be/X-ray binary system,
and to neutron stars having long spin periods.

\subsection{Comparison with PSR B1259$-$63} 

There are three binary radio pulsars known to have non-degenerate
orbital companions: PSRs J0045$-$7319 \citep{kjb+94}, J1740$-$3052
\citep{sml+01}, and B1259$-$63 \citep{jml+92}. Of these three, only
PSR B1259$-$63 orbits a Be star. There are also many Be/X-ray HMXBs
that are known, including more than 60 in the SMC \citep{hp08}, but
apart from PSR B1259$-$63, none of these has yet been observed as a
radio pulsar.  For this reason, PSR B1259$-$63 is a natural system
with which to compare our results for XTE~J0103$-$728.

PSR B1259$-$63 is in a highly eccentric and wide orbit, with an
eccentricity of 0.87 and an orbital period of 3.4 yr \citep{jml+94}
The pulsar passes quite close to its Be star companion at periastron,
coming within 24 stellar radii of the star \citep{jbw+05}. This
distance is comparable to the estimated radius of the circumstellar
disk of the Be star, which is thought to be at least 20 stellar radii
in size \citep{jml+94}.  Near periastron, the radio pulses from PSR
B1259$-$63 are eclipsed for $\sim 40$ days by the circumstellar disk.
Enhanced X-ray emission is also detected near periastron, since the
pulsar interacts with the stellar wind \citep{ktn+95}. However, pulsed
X-rays have not been detected from PSR B1259$-$63, unlike
XTE~J0103$-$728.  \citet{mjm95} modeled the B1259$-$63 system near
periastron and investigated two models which could explain the
observed eclipsing behavior. They found that a model employing a cool,
dense equatorial disk around the Be star successfully predicted the
observed eclipse duration at $\sim 1.5$ GHz, if free-free absorption
were the primary eclipsing mechanism. They also found that scattering
in the disk may be contributing to the radio eclipse.  In any case, it
seems reasonable that free-free absorption and scattering could also
be preventing the detection of radio pulses from XTE~J0103$-$728 if it
is in a similarly eccentric orbit and was near periastron at the time
of our observation.

The 1400 MHz flux density of pulsed emission from PSR B1259$-$63 has
been measured and published in a variety of papers using data
collected over the course of several orbits
\citep{hfs+04,jml+92,jml+94,jml+96,jbw+05,cjm+02}. The measured values
vary considerably, but consideration of all the measurements suggests
an average flux density of $\sim 5$ mJy around 1400 MHz. We use this
estimate for the comparison with our observation of
XTE~J0103$-$728. Near periastron, when PSR B1259$-$63 is eclipsed,
there is no detectable radio emission at this frequency.  We adopt
$\sim 1$ mJy as a conservative upper limit on the flux density during
eclipse by considering the uncertainties in the 1400 MHz flux reported
both before and after the eclipse during the 2000 and 2004 periastron
passages \citep{cjm+02,jbw+05}.  The best estimate for the distance to
PSR B1259$-$63 is 1.5 kpc, based on optical observations of the Be
star companion \citep{jml+94}. From this, the resulting 1400 MHz
luminosities are calculated to be $\sim 11$ mJy kpc$^{2}$ away from
periastron and $\la 2$ mJy kpc$^{2}$ during eclipse near
periastron. In either case, the luminosity is below the limit of $\sim
45$ mJy kpc$^{2}$ that we have established for pulsed emission from
XTE~J0103$-$728 at the epoch of our observation. We therefore cannot
rule out radio emission from XTE~J0103$-$728 that is comparable in
luminosity to the emission seen from PSR B1259$-$63.

\subsection{Comparison with Other Long-period Neutron Stars} 

In this section, we compare the luminosity upper limit for
XTE~J0103$-$728 to the radio luminosities (or luminosity upper limits
in some cases) of neutron stars which have long spin periods like
XTE~J0103$-$728.

In the ATNF pulsar catalog \citep{mht+05}, there are five radio
pulsars with periods greater than 6 s that have measured 1400 MHz flux
densities and corresponding estimated luminosities. The radio
luminosities of these pulsars range from 0.03 to 23 mJy kpc$^{2}$,
which is below our upper limit of $\sim 45$ mJy kpc$^{2}$ for periodic emission
from XTE~J0103$-$728.  The sample of six long-period XDINs for which
new, sensitive radio limits have recently been obtained using the
Green Bank Telescope \citep{kml+08} have 1400-MHz luminosity limits
that are several orders of magnitude below our limit. These XDINs have
periods ranging from 7 to 11 seconds.  There are five AXPs with
periods longer than 6 s which have been searched at 1400 MHz for radio
emission and for which luminosity upper limits have been established
\citep{bri+06,chk07,bri+06a}. Although these AXPs are believed to have
much stronger magnetic fields than the neutron stars in these other
classes, they may have similar beaming and detectability selection
effects in the radio owing to their similarly long spin periods (see
discussion below).  The luminosity limits established for these AXPs
are all around 1 mJy kpc$^{2}$, which is more than an order of
magnitude below our limit for XTE~J0103$-$728. A radio search was also
previously conducted on SGR~0526$-$66 in the Large Magellanic Cloud,
which, like the AXPs, is also believed to be a magnetar.  No radio
emission was detected from SGR~0526$-$66 down to an upper limit of
$\sim 350$ mJy kpc$^{2}$ at 1400 MHz \citep{cpk+02}.

By looking at these values, one can see that it is entirely possible
that XTE~J0103$-$728 could have radio emission comparable in strength
to these long period radio pulsars, XDINs, and AXPs (if these latter
two classes are in fact radio emitters), and that we are simply not
detecting radio pulsations due to a lack of
sensitivity. Unfortunately, the large distance to XTE~J0103$-$728
makes the limit impossible to significantly improve upon until
next-generation instruments, such as the Australia Square Kilometer
Array Pathfinder \citep[ASKAP;][]{j++07,j++08}, can be developed and
used. Even if XTE~J0103$-$728 were a strong radio emitter,
its radio beam might not be aligned with its X-ray beam and could miss
our line of sight.  If this were the case, XTE J0103$-$728 might
remain undetectable to us in the radio owing to its emission geometry.

Narrow pulse widths are also generally observed for long-period radio
pulsars. For instance, the two cataloged radio pulsars with the
longest spin periods, PSRs J2144$-$3933 and J1001$-$5939, have
full-width half maximum radio pulse widths of 0.2\% and 0.5\% of the
pulse period, respectively \citep{ym99,lfl+06}. These duty cycles are
among the smallest measured for the radio pulsar population and
indicate the severity of the selection effects that might be
preventing the detection of many long-period radio pulsars, of which
XTE~J0103$-$728 might be one.  Note that the small duty cycle
associated with the narrow pulses would partially compensate for the
sensitivity selection effect on long-period pulsars 
from the high-pass filtering (described
above), since there would be many high-frequency harmonics in the
signal.

\section{Conclusions} 

The X-ray pulsations from XTE~J0103$-$728 that were first detected by
\citet{cmm+03} with RXTE were detectable over the course of a few
weeks. This suggests that the source underwent a Type II outburst,
caused by the expansion of the circumstellar disk around the Be star,
with subsequent enhanced accretion onto the neutron star surface
\citep{hep08,hp08}.  If, however, the X-ray emission detected in 2006
was from a Type I burst, which lasts for a few days during periastron
passage and recurs periodically with the orbital period of the system
\citep[e.g.,][]{on01}, then our radio observation likely occurred well
away from periastron, assuming the orbit of XTE~J0103$-$728 is
sufficiently wide and eccentric, as is the case for most Be/X-ray
binaries \citep{lvv00}.  In this case, radio pulses from
XTE~J0103$-$728 may be more likely to be seen, since radio pulses from
PSR B1259$-$63 are detectable for the majority of its 3.4 yr orbit.

This also points to the value of repeated radio search observations of
Be/X-ray binaries such as XTE J0103$-$728, since their emission
properties can vary significantly at different orbital phases.  Future
radio search observations of XTE J0103$-$728 would be valuable,
especially if observations can be conducted that could sample the
orbital phase of the system well.  It may also be the case that
XTE~J0103$-$728 emits only very weak radio emission (or perhaps no
radio emission at all), that the radio and X-ray beams are not aligned
with each other, or that a strong but narrow radio beam misses our
line of sight, making this source undetectable in the radio.

\acknowledgements

We thank John Reynolds for the allocation of director's time for this
search observation at Parkes and the Parkes Observatory staff for
hospitality and assistance with observations. We also thank the
referee for a careful reading of the manuscript and for a number of
helpful suggestions which have significantly improved the discussion
of the results. The Parkes radio telescope is part of the Australia
Telescope, which is funded by the Commonwealth of Australia for
operation as a National Facility managed by CSIRO. FC was supported by
grants from Research Corporation, the Mount Cuba Astronomical
Foundation, and the National Radio Astronomy Observatory. DRL and VIK
acknowledge support from West Virginia EPSCoR in the form of a
Research Challenge Award.

\begin{deluxetable}{lc}
\tablecaption{1400 MHz Radio Limits for XTE~J0103$-$728.\label{tbl-1}}
\tablewidth{0pt}
\tablehead{
\colhead{Parameter} &
\colhead{Value} 
}
\startdata
Observation date & 1 Feb 2008 \\
Observation MJD   & 54497 \\
1400 MHz flux density limit (periodic emission)\tablenotemark{a}  & $13 \mu$Jy \\ 
1400 MHz luminosity limit (periodic emission)\tablenotemark{b}    & 45 mJy kpc$^{2}$ \\
1400 MHz flux density limit (single pulses)\tablenotemark{c}  & 0.60-0.03 Jy \\
1400 MHz luminosity limit (single pulses)\tablenotemark{b}    & 2150-120 Jy kpc$^{2}$  
\enddata


\tablenotetext{a}{~Limit on periodic emission obtained from the
modified radiometer equation assuming a 5\% pulsed duty cycle and no
degradation in sensitivity from RFI.  The limit obtained from the
blind Fourier search is roughly a factor of two higher (less
sensitive) owing to the different detection algorithms used.  The
limits become significantly higher (less sensitive) than those
calculated with the modified radiometer equation if a large duty cycle
is assumed ($\ga 20$-30\%). This is due to the high-pass filtering
present in the observing system which reduces sensitivity to pulse
widths $\ga 1$~s.}

\tablenotetext{b}{~Luminosity limit obtained assuming a distance of
$\sim 60$~kpc, the distance to the SMC \citep{hhh05}.}

\tablenotetext{c}{~Limits on single pulse emission for boxcar filter
widths ranging from 0.5 ms to 1024 ms. The limits were obtained using
the radiometer equation with an integration time equal to the window
width and a detection S/N threshold that changed from 8 to 20 for
filter widths greater than or equal to 4 ms. This change in detection
threshold was chosen to account for the increased presence of RFI in
the larger filter windows.}

\end{deluxetable}

\end{document}